\documentclass{article}

\usepackage{arxiv}

\usepackage[utf8]{inputenc} % allow utf-8 input
\usepackage[T1]{fontenc}    % use 8-bit T1 fonts
\usepackage{hyperref}       % hyperlinks
\usepackage{url}            % simple URL typesetting
\usepackage{booktabs}       % professional-quality tables
\usepackage{amsfonts}       % blackboard math symbols
\usepackage{nicefrac}       % compact symbols for 1/2, etc.
\usepackage{microtype}      % microtypography
\usepackage{lipsum}		% Can be removed after putting your text content
\usepackage{graphicx}
\usepackage{natbib}
\usepackage{doi}
\usepackage{amsmath}

\title{'Guided' Fractures in Graphene Mechanical Diode-like Structures}

\author{ %\href{https://orcid.org/0000-0000-0000-0000}{\includegraphics[scale=0.06]{orcid.pdf}
    \hspace{1mm}Levi C.~Felix and Douglas S.~Galvao\thanks{Corresponding author: galvao@ifi.unicamp.br} \\
	Department of Applied Physics\\
	State University of Campinas\\
	Campinas, SP 13083-970, Brazil \\
}

\renewcommand{\shorttitle}{Graphene fracture diode}

\hypersetup{
pdftitle={A template for the arxiv style},
pdfsubject={q-bio.NC, q-bio.QM},
pdfauthor={David S.~Hippocampus, Elias D.~Striatum},
pdfkeywords={First keyword, Second keyword, More},
}

\begin{document}
\maketitle

\begin{abstract}
	The concept of diode is usually applied to electronic and thermal devices but very rarely for mechanical ones. A recent proposed fracture rectification effect in polymer-based structures with triangular voids defects has motivated us to test these ideas at the nanoscale using graphene membranes. Using fully-atomistic reactive molecular dynamics simulations we showed that a robust rectification-like effects exists. The fracture can be 'guided' to easier propagate along one specific direction than its opposite. We also observed that there is an optimal value for the spacing between each void for rectification effect. 
\end{abstract}

% keywords can be removed

\section{Introduction}

Graphene has received great attention from the scientific community due to its unique electronic properties~\cite{novoselov_2004,geim_2007,zhang_2005,young_2009,castroneto_2009}. However, graphene is also interesting due to its excellent mechanical properties~\cite{lee_2008,zhang_2005,bizao_2017,bizao_2018,fonseca_2021}. It is believed to be one of the toughest materials (natural and synthesized), with Young's modulus of $\sim 1$ TPa~\cite{lee_2008,lee_2012}, resulting from its strong $\sigma$ bonds originating from the sp$^2$ carbon hybridization. It also presents nonlinear elastic behavior~\cite{lee_2008,branicio_2016} that leads to a maximum stress value that characterizes its intrinsic strength. Besides the elastic properties, the fracture dynamics of graphene have been also experimentally investigated~\cite{zhang_2014} through nanoindentation. An important property that characterizes fracture in materials is the fracture toughness ($K_{C}$), which describes the resistance to fracture of a material containing a crack tip. For graphene, a value of $K_{C} \sim 4.0$ MPa~m$^{1/2}$ was measured. Typical values of $K_{C}$ are 62-280 MPa~m$^{1/2}$ for stainless steels, 22-35 MPa~m$^{1/2}$ for aluminum alloys and 1.19-4.30 MPa~m$^{1/2}$ for ABS polymers~\cite{ashby_2011}. Although, $K_C$ for graphene is not as high as the values just mentioned, its specific mass is much lower than the materials previously mentioned. Thus, graphene is an outstanding candidate for applications in composite structures that require simultaneously lightweight and high strength materials~\cite{young_2012,kim_2013,yang_2013}. Since mechanical failure is sometimes unavoidable, fracture properties should be taken into account when designing new materials for mechanical and structural applications~\cite{ashby_2011}.

Many composite structures, such as fiber-reinforced ones, possess anisotropic properties. Their stiffness and fracture toughness values along the direction of the fibers are very distinct from the transverse ones~\cite{callisterjr_2018}. Some nanomaterials, like 2D black phosphorus, also possess anisotropic mechanical behavior in their two distinct in-plane directions: armchair and zigzag~\cite{sha_2015,liu_2019}. However, the centrosymmetry of these systems does not allow any property to change with a reversal of direction. There are systems that do possess properties that are dependent on the direction and also on the propagation pathways, but they rely on the presence of interfacial effects~\cite{daniel_2004,malvadkar_2010,hancock_2012,zambrano_2018,xia_2012}. Recent works have demonstrated rectification effects and deflection of fractures by modifying only the microstructure with the creation of structural voids~\cite{hossain_2014,brodnik_2021,conway_2021} without any introduction of a different material. These strategies serve as a way to mitigate the damage of critical components when a fracture is unavoidable~\cite{brodnik_2021}.

The concept of diode is usually applied to electronic and thermal devices but very rarely for mechanical ones. Recently, Brodnik \textit{et al}~\cite{brodnik_2021} reported fracture diode-like structures by creating structural voids arrangements in millimeter-sized metamaterials. A natural question is whether this behavior still holds at the nanoscale and for the reasons discussed above, graphene membranes would be a good choice. In this work, we further explored these ideas by creating triangular voids into a graphene sheet (see Figure~\ref{fig:scheme}) in order to test the possibility of having asymmetrical fracture behavior. First, we considered a structure with two notches at the left and right ends to determine the preferential crack propagation direction. Once determined this direction, we tested the diode-like behavior considering just only one notch to determine which direction the fractures experience less resistance (tensile strength in the stress-strain curve). Although there are several works on electronic~\cite{dragoman_2010,dibartolomeo_2016,wang_2021,kim_2013a,li_2016} and thermal~\cite{wang_2017,liu_2021,chen_2020,melis_2015,wang_2014} graphene-based diodes, to our knowledge mechanical graphene-based diode based on ideas proposed by Brodnik \textit{et al}~\cite{brodnik_2021} has not been investigated and it is one of the goals of the present study.

\section{Methodology}

\begin{figure}[!htb]
\centering
\includegraphics[width=\linewidth]{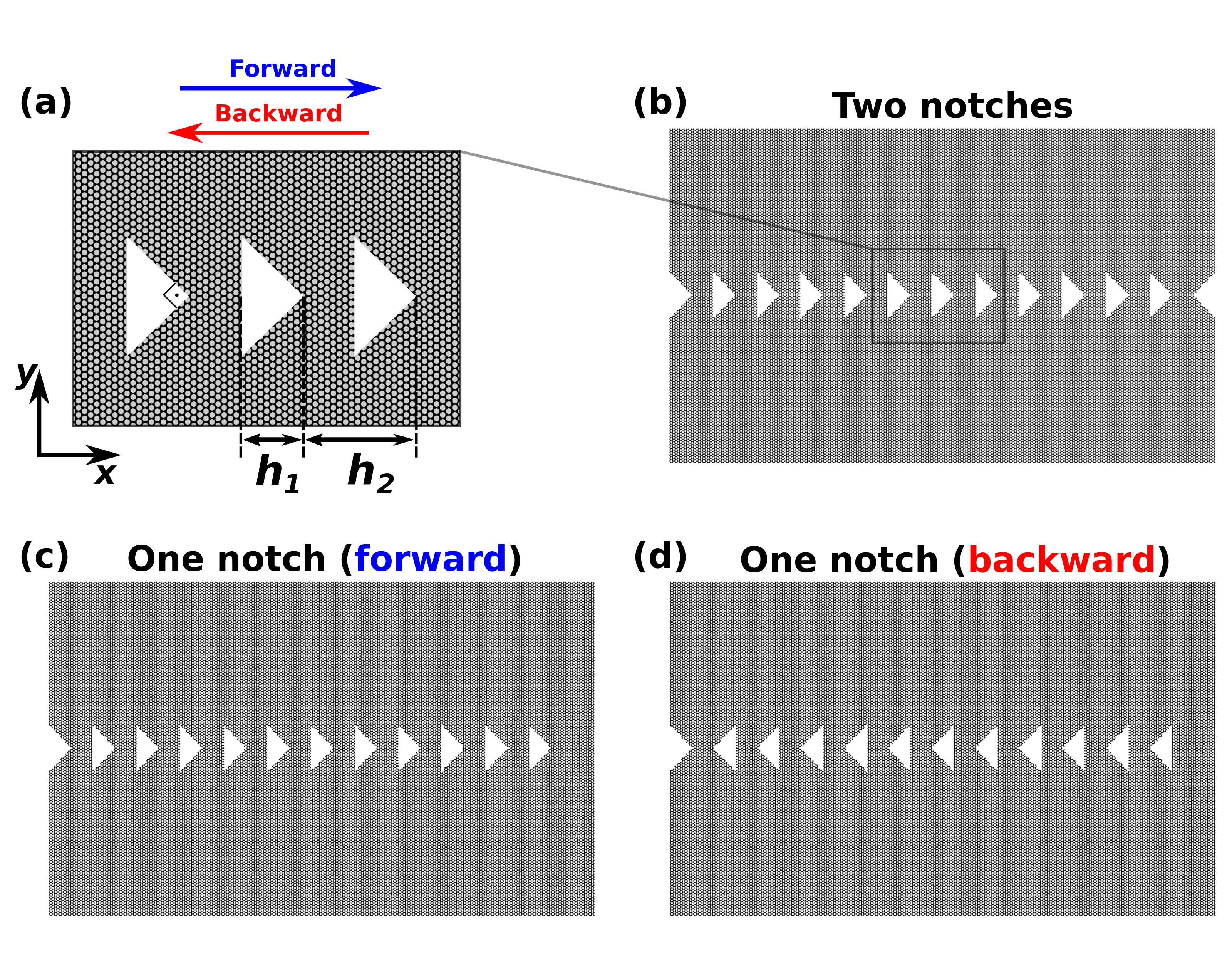}
\caption{Scheme of the structural models investigated here. (a) Forward/backward direction definition along with the triangle dimensions. (b) The two-notch structure, which a fracture could possibly start. One-notch structures corresponding to the (c) forward and (d) backward directions, respectively.}
\label{fig:scheme}
\end{figure}

From a pristine graphene sheet with dimensions of $L_x=30$ nm and $L_y=50$ nm, we created holes along its middle line. These holes are isosceles triangles with $h_1=2$~nm of height and are equally translated along the $x$ direction by $h_2$ (see Figure \ref{fig:scheme}~(a)). We considered two cases where (i) two triangular notches at the left and right edges are present as crack tips to start the fracture propagation as shown in Figure \ref{fig:scheme}~(b) and (ii) with only one notch to induce a fracture to propagate along the direction where the triangles are pointing (here denoted as forward direction, as shown in Figure \ref{fig:scheme}~(a)) and through the opposite direction (backward, as also illustrated in Figure \ref{fig:scheme}~(a)), as illustrated in Figure \ref{fig:scheme}~(c) and (d). The latter two configurations were considered to determine the preferential fracture propagation direction (which one offers lesser resistance to crack propagation), here estimated by their tensile strength. Additionally, we also compared the cases with different $h_2$ translations. The stretching direction was chosen to be along the $y$ axis and, thus, a periodic boundary condition was applied to this direction. A vacuum region of 3 nm is introduced along $x$ and $z$ directions to mimic a finite nanoribbon. The tensile deformations were investigated through fully atomistic Molecular Dynamics (MD) simulations using the open-source code LAMMPS~\cite{plimpton_1995}. The interatomic interaction among the carbon atoms was modeled with the reactive AIREBO potential~\cite{stuart_2000}. To better describe the mechanical behavior at large strain values, we modified the minimum cutoff radius of the force field to 2~\AA~as described by Shenderova \textit{et al}~\cite{shenderova_2000}. Before stretching, an NPT equilibration MD runs, at a temperature of $T = 300$~K, was performed during $100$~ps to eliminate any residual stresses in the structures. During NPT MD runs, only the $y$ dimension was allowed to vary, since it is the only periodic one. The structural stretching process is implemented by increasing the length of the simulation box along the $y$ direction with a constant engineering strain rate of $\gamma = 10^{-5}$ fs$^{-1}$ and an MD timestep of $dt = 0.1$ fs. In this way, the $y$ dimension of the simulation box vary as $L_y(t) = L_y(0) (1 + \gamma  t)$. The deformation level of all structures studied here is quantified by the engineering strain $\varepsilon = (L_y(t) - L_y(0))/L_y(0)$.

To obtain the stress-strain response, we calculated the total stress, given by
\begin{equation}
    \sigma_{ab} = \frac{1}{V}\sum^N_{k=1} m_k v_{k,a}v_{k,b} + \frac{1}{V}\sum^N_{k=1}r_{k,a}f_{k,b},
\end{equation}
where the indices $a,b = x,y,z$ correspond to the Cartesian directions, $N$ is the number of atoms in the system, $m_k$ is the atomic mass of the $k$-th atom, $r_{k,a}$ ($v_{k,a}$) is the atomic position (velocity) of the $k$-th along the $a$ direction, $f_{k,b}$ is the corresponding force on the $k$-th atom due to its neighbors. The volume of the structure is given by $V=L_x~L_y~L_z$, where a thickness of $L_z=3.35$~\AA~ is adopted for graphene.

To obtain further insights on the local stress accumulation and fracture dynamics, we also computed the virial stress per atom $k$, defined by
\begin{equation}\label{eq:atomicvirial}
    \begin{split}
        \sigma^k_{ab} &= r_{k,a}f_{k,b}= \frac{1}{2}\sum_{l = 1}^{N_p} (r_{k,a} F^{(2)}_{k,b} + r_{l,a} F^{(2)}_{l,b}) \\ &+ \frac{1}{3}\sum_{m = 1}^{N_p}\sum_{n = 1}^{N_p} (r_{k,a} F^{(3)}_{k,b} + r_{m,a} F^{(3)}_{m,b} + r_{n,a} F^{(3)}_{n,b}) + \sum_{p = 1}^{N_f} r_{k,a} \mathcal{F}^{(p)}_{k,b},
    \end{split}
\end{equation}
where $n$ runs over the $N_p$ neighbors of atom $k$, $F^{(2)}_{k,b}$ is the $b$ pairwise interaction force component on atom $k$, $m,n$ take into consideration all pair atoms in the three-body $F^{(3)}_{k,b}$ force terms on the $k$-th atom, $p$ runs over all LAMMPS fixes that apply internal constraint forces to atom $k$, and $\mathcal{F}^{(m)}_{k,b}$ is the corresponding force due to $p$-th fix. Due to the von Mises failure criteria, it is more effective to analyze the local distribution of stress with the atomic von Mises stress~\cite{mises_1913} given by the relation
\begin{equation}\label{eq:atomvonmises}
    \sigma^{k}_{v} = \sqrt{\frac{(\sigma^{k}_{xx} - \sigma^{k}_{yy})^2 + (\sigma^{k}_{yy} - \sigma^{k}_{zz})^2 + (\sigma^{k}_{xx} - \sigma^{k}_{zz})^2 + 6((\sigma^k_{xy})^2+(\sigma^k_{yz})^2+(\sigma^k_{zx})^2)}{2}}.
\end{equation}
To better visualize stress accumulations, the values of $\sigma_v^k$ for each atom are taken as the average between its corresponding neighbors within a cutoff distance of $10$~\AA~using OVITO~\cite{stukowski_2010}.

\section{Results}

The forward and backward directions are defined in Figure \ref{fig:scheme}~(a). We first performed the stretching simulations on the two-notch structures with different void spacing $h_2$ in order to determine the preferential fracture propagation direction (left or right).

\begin{figure}[!htb]
\centering
\includegraphics[width=\linewidth]{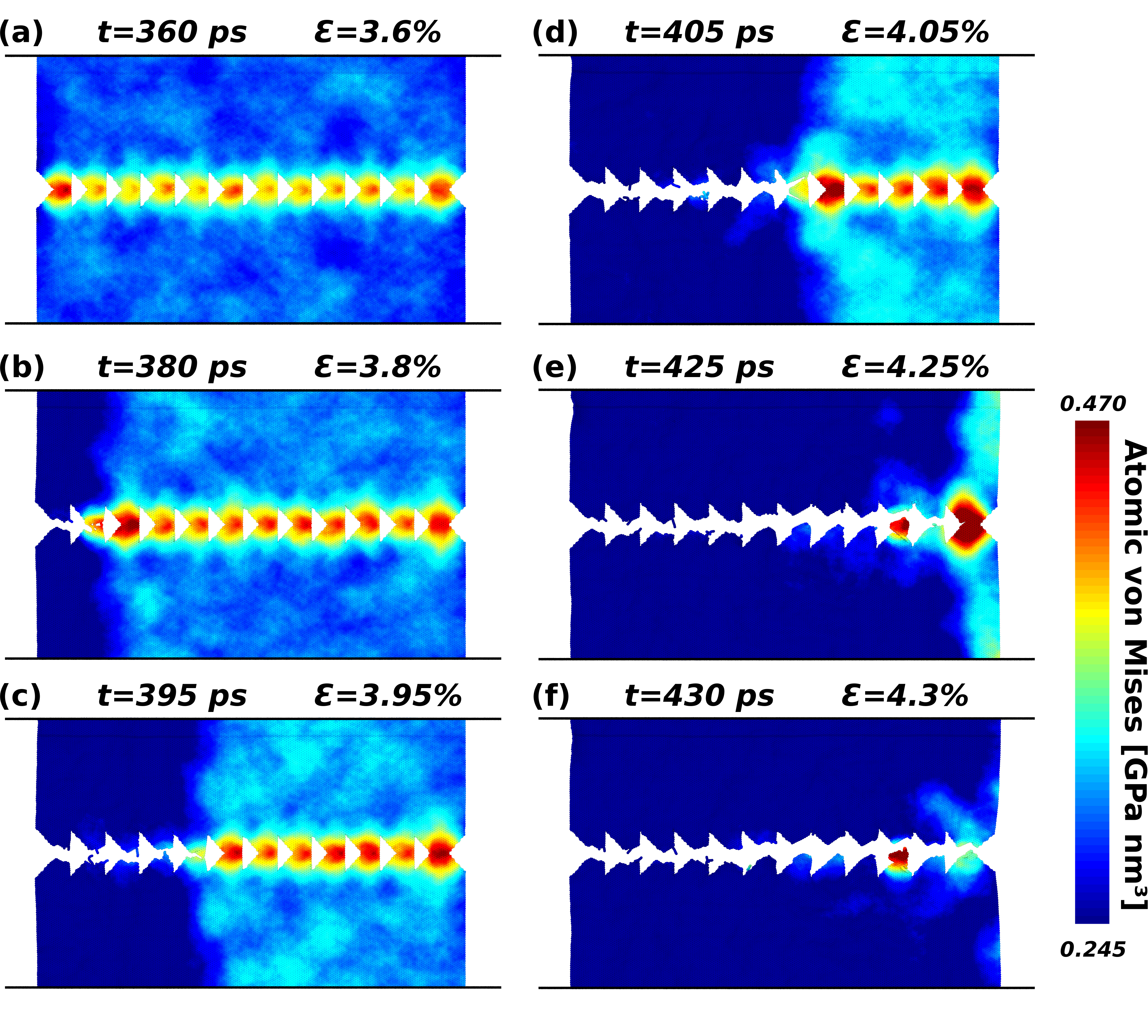}
\caption{Crack propagation for a two-notch structure for the case of $h_2/h_1=2$. (a)-(f) Representative MD snapshots at successive instants of the simulations. The time $t$ is given in picoseconds and the corresponding tensile strain $\varepsilon = \gamma t$ in percentage values. Atom colors indicate atomic von Mises stress values according to the color bar located on the right. Blue and red colors indicate low and high-stress values, respectively.}
\label{fig:uni-2-notch-2}
\end{figure}

In Figure \ref{fig:uni-2-notch-2}, we present representative MD snapshots at different simulation times and their corresponding spatial stress distribution. It is clear that once started the fracture propagates preferentially from left to right corresponding to the forward direction. From the von Mises distribution, we can also see that, as expected, the stress is released with the fracture, which can be seen by the blue region that follows the crack as the structure breaks. The stress peaks, shown in Figure \ref{fig:uni-2-notch-2} as red spots between voids, are very spatially localized where the larger spots on the beginning are situated on the left and right ends due to the presence of the crack tips. The detailed fracture propagation is shown in the supplementary video S1.

\begin{figure}[!htb]
\centering
\includegraphics[width=\linewidth]{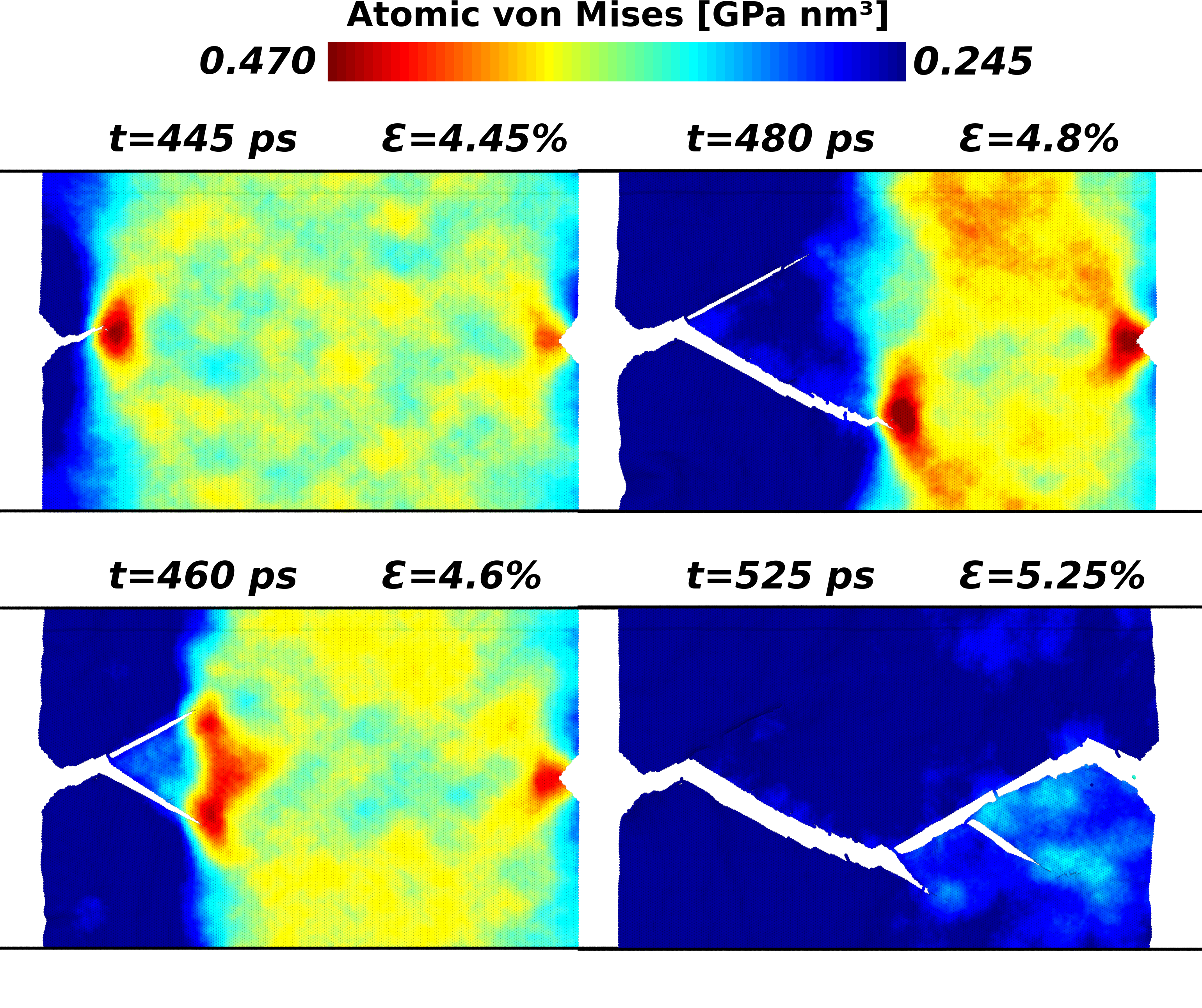}
\caption{Crack propagation in a two-notch structure for the case with no internal voids. Representative MD snapshots at successive instants of the simulations. The time $t$ is given in picoseconds and the corresponding tensile strain $\varepsilon = \gamma t$ in percentage values. Atom colors indicate atomic von Mises stress values according to the color bar located on the right. Blue and red colors indicate low and high-stress values, respectively.}
\label{fig:uni-2-notch-pristine}
\end{figure}

From Figure \ref{fig:uni-2-notch-2} we observed that the fracture propagated along the array of triangular voids in the middle line of the structure. In order to determine whether this happens due to the presence of the defects, we run a similar MD simulation with two notches but without the internal triangular voids. As we can see in Figure \ref{fig:uni-2-notch-pristine}, the crack starts and propagates from the left by it rapidly deviates from the middle line following specific angles. These angles (the so-called 'petal' angles) are characteristic of fractured hexagonal lattices~\cite{bizao_2017}. Therefore, the array of triangular voids serves as a structural mechanism to 'guide' the fracture to a specific region of the structure, in our case the middle line.

\begin{figure}[!htb]
\centering
\includegraphics[width=\linewidth]{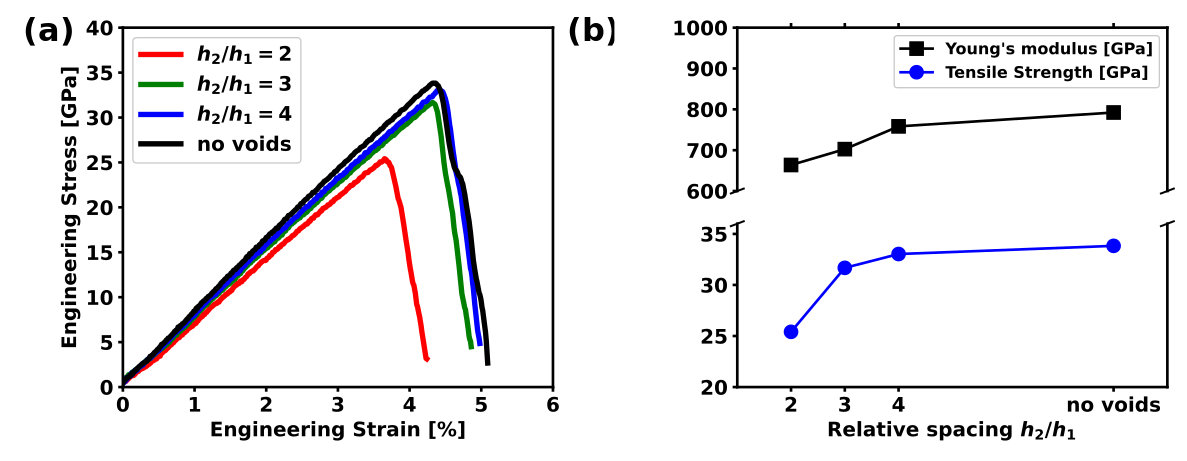}
\caption{Stress-strain curves of the two-notch structures as a function of the $h_2/h_1$ ratios. (a) Engineering stress-strain curve for different spacing between the triangular voids, and; (b) The corresponding Young's modulus and tensile strength values.}
\label{fig:stress-strain-2-notch}
\end{figure}

We also carried out stretching simulations for the cases of void spacing with $h_2 = 3h_1$ and $h_2 = 4h_1$. The corresponding stress-strain curves are shown in Figure \ref{fig:stress-strain-2-notch}~(a). A common behavior is observed for all cases: (i) a linear regime at low strain values (up to $\varepsilon \sim 1\%$) characteristic of an elastic behavior; (ii) a slight increase of the slope corresponding to the propagation of fracture from one void to another, where more pronounced stress peaks appear near the crack tip, and; (iii) a maximum peak indicating the tensile strength of a given structure that is subsequently followed by an abrupt drop of stress values, which indicates total fracture. As the relative spacing $h_2$ increases, both tensile strength and Young's modulus (here estimated as the slope of the linear regime) also increases. This behavior is expected since the stiffness of graphene decreases with increased defect density. These trends are clearly shown in Figure \ref{fig:stress-strain-2-notch}~(b).

\begin{figure}[!htb]
\centering
\includegraphics[width=\linewidth]{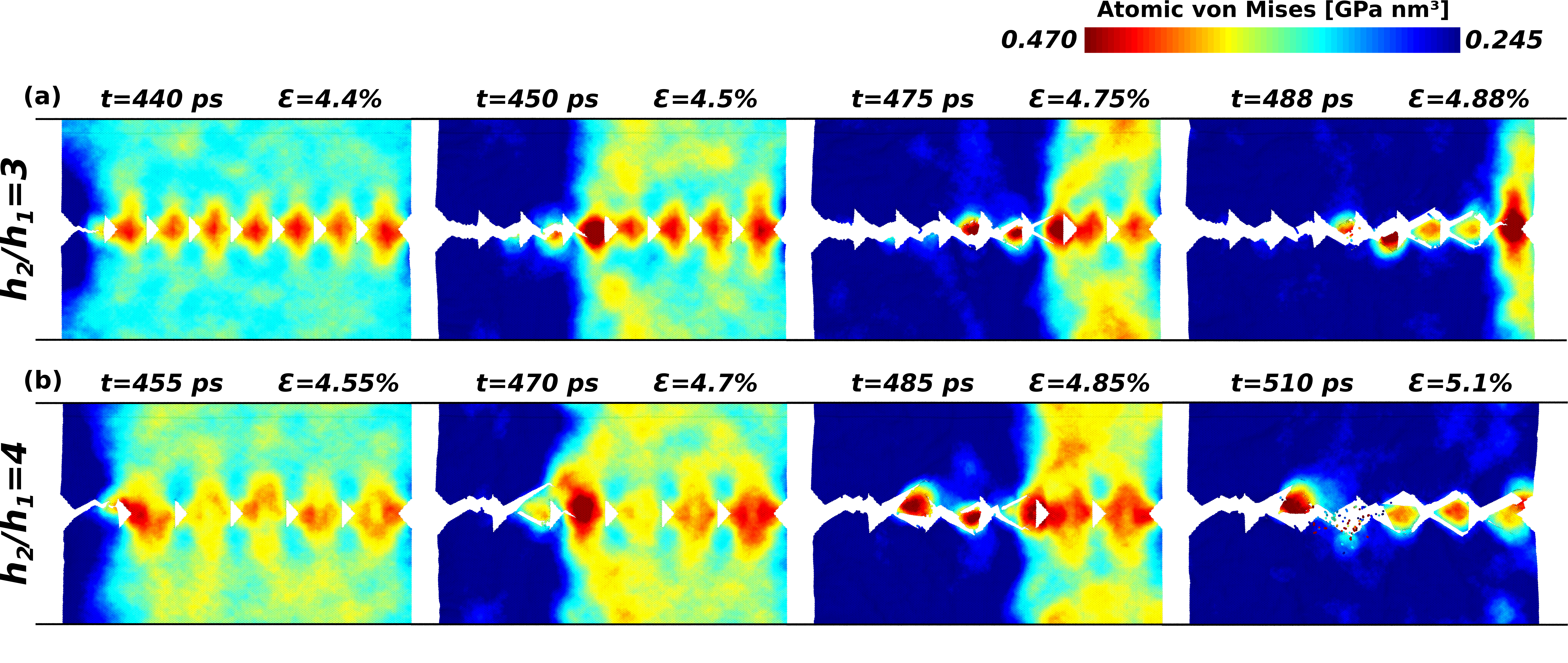}
\caption{Representative MD snapshots at successive instants of crack propagation in a two-notch structure for the cases (a) $h_2/h_1=3$, and (b) $h_2/h_1=4$.  The time $t$ is given in picoseconds and the corresponding tensile strain $\varepsilon = \gamma t$ in percentage values. Atom colors indicate atomic von Mises stress values according to the color bar located on the right. Blue and red colors indicate low and high-stress values, respectively.}
\label{fig:uni-2-notch-3-4}
\end{figure}

The 'guided' fracture effect is also present in cases where $h_2 = 3h_1$ and $h_2 = 4h_1$, as shown in Figure \ref{fig:uni-2-notch-3-4}~(a) and (b), respectively. From the time evolution indicated above each snapshot, fracture propagates slower for higher $h_2$ values. This occurs because each void accelerates the process where there is nothing opposing to crack propagation, whereas in high-$h_2$ structures there is more material to break, which delays the total fracture.

\begin{figure}[!htb]
\centering
\includegraphics[width=\linewidth]{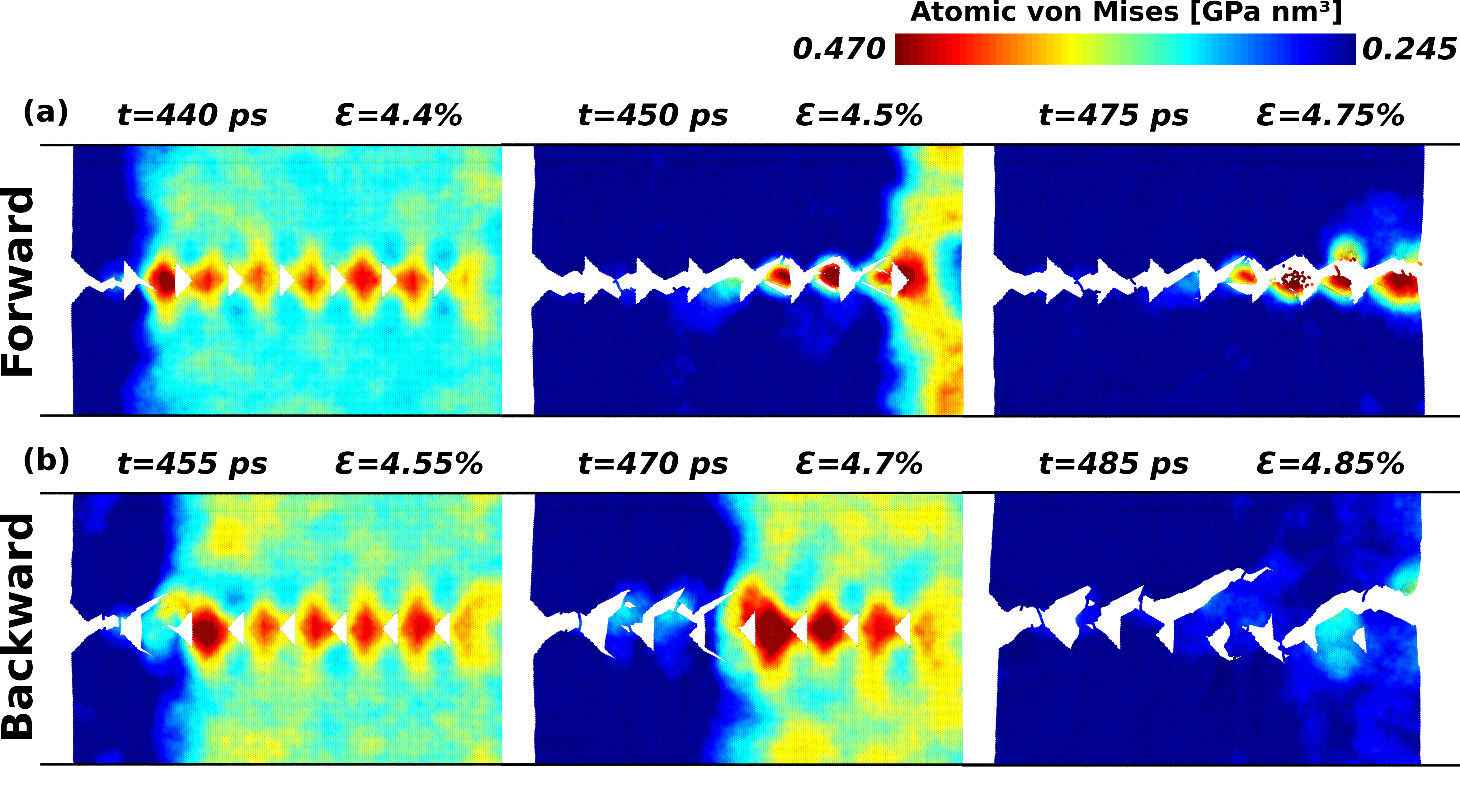}
\caption{Representative MD snapshots at successive instants of the crack propagation in a one-notch structure for the case where $h_2/h_1=3$. The time $t$ is given in picoseconds and the corresponding tensile strain $\varepsilon = \gamma t$ in percentage values. Atom colors indicate atomic von Mises stress values according to the color bar located on the right. Blue and red colors indicate low and high-stress values, respectively. (a) and (b) show the fracture process along the forward and backward directions, respectively.}
\label{fig:uni-1-notch-f-b}
\end{figure}

Once determined the preferential fracture direction propagation, we proceed to test the diode-like behavior using the same configurations of triangular voids with a spacing of $h_2$ but just with only one notch at the left. This favors the fracture to propagate along a specific direction, as discussed above. The temporal evolution of the crack propagation along the forward and backward directions are presented in Figure\ref{fig:uni-1-notch-f-b}~(a) and (b), respectively. Along the forward direction, the fracture is guided by the array of voids as in the two-notch structure (Figure \ref{fig:uni-2-notch-3-4}~(a)), where only carbon bonds near the middle line are broken in the fracture process. By constraining the crack to propagate along the backward direction (inverting the triangular voids orientation), the fracture also occurs but propagating away from the voids and, therefore, the 'guided' fracture effect is no longer present. This indicates that such an array of triangular voids creates a mechanical diode-like effect for the graphene fracture at the nanoscale, similarly to tailored voids in polymeric materials at the macroscale~\cite{brodnik_2021}, thus confirming that the effect still holds at the nanoscale. The detailed fracture propagation of the forward and backward cases are shown in the supplementary videos S2 and S3, respectively.

\begin{figure}[!htb]
\centering
\includegraphics[width=\linewidth]{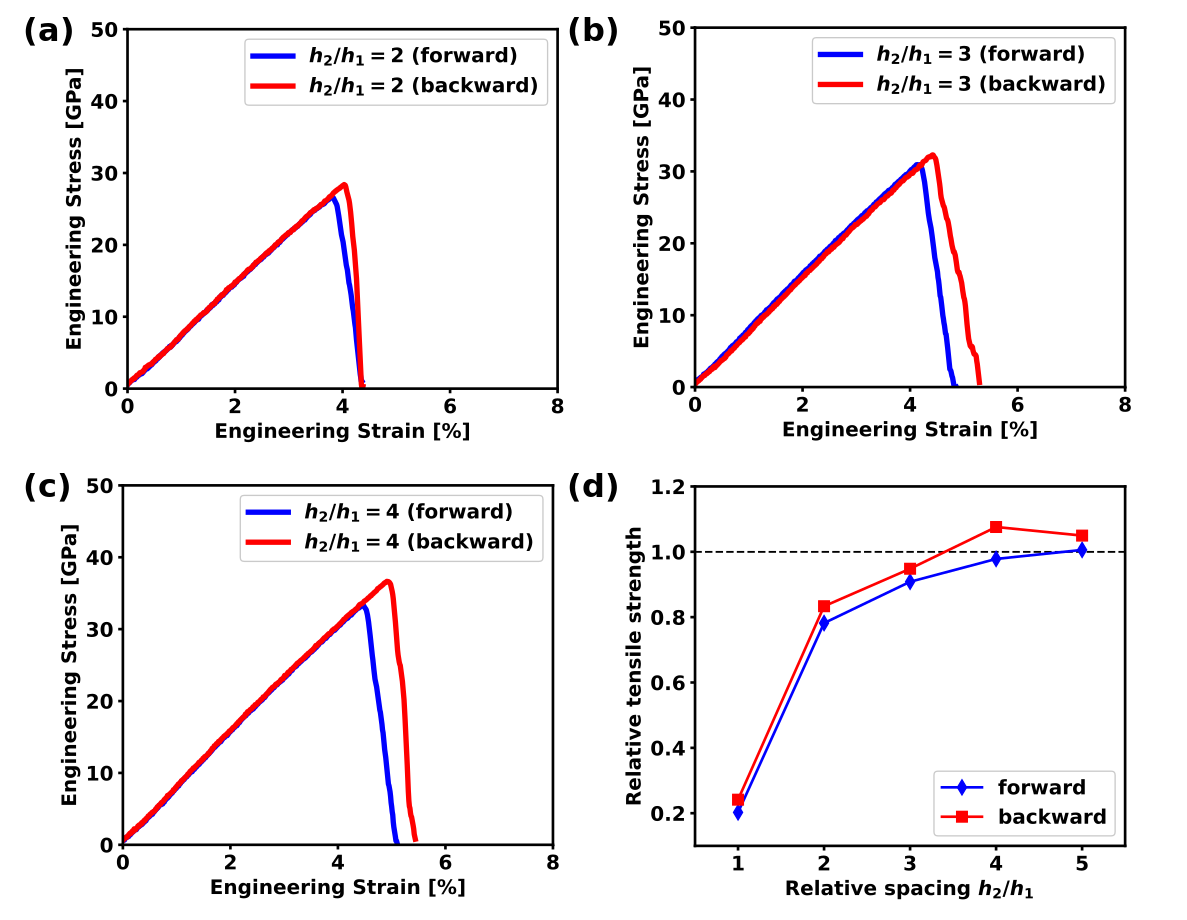}
\caption{Stress-strain curves for the forward and backward directions for (a) $h_2/h_1=2$, (b) $h_2/h_1=3$, and (c) $h_2/h_1=4$. (d) Asymmetry of tensile strength values for the forward and backward direction as a function of relative spacing $h_2/h_1$ between each triangular void.}
\label{fig:stress-strain-f-b}
\end{figure}

The robustness of fracture 'rectification' provided by such an array of defects is quantified in Figure \ref{fig:stress-strain-f-b}. For a given void spacing, Young's modulus values are not considerably affected by the change in the crack propagation direction, since the stress values are essentially the same for low strain values as can be seen in Figure \ref{fig:stress-strain-f-b}~(a)-(c). From this Figure, we can also see that the rectification effect increases with $h_2$ up to $4h_1$. This trend can be explained based on that more bonds far from the middle line are broken in the propagation of fracture where larger distances between successive voids mean more energy needed for the fracture to propagate from one void to another. However, this trend in relative tensile strength should have an upper limit because the increase of $h_2$ values reaches a limit where two neighbor triangles are very apart and the system as a whole resembles the case without any voids or, in other words, becomes symmetrical.

%\begin{figure}[!htb]
%\centering
%\includegraphics[width=\linewidth]{mises-x.png}
%\caption{Scatter plot of the atomic von Mises stress distribution along the $x$ position in the forward and backward directions for cases where (a) $h_2/h_1=2$, (b) $h_2/h_1=3$, and (c) $h_2/h_1=4$. Each point represents an atom and its $x$ ($y$) value correspond to its position along the $x$-axis (atomic von Mises stress). }
%\label{fig:mises-x}
%\end{figure}

%\newcolumntype{C}{>{\centering\arraybackslash}X}
%\begin{table}[!htb]
%\centering
%\caption{Insert caption here.}
%\begin{tabularx}{1.0\textwidth}{ C | C | C | C }
%\toprule
%Case & Method\#1 & Method\#2 & Method\#3 \\ 
%\midrule
%1 & 50 & 837 & 970 \\
%2 & 47 & 877 & 230 \\
%3 & 31 & 25 & 415 \\
%4 & 35 & 144 & 2356 \\
%5 & 45 & 300 & 556 \\
%\bottomrule
%\end{tabularx}
%\label{table:tab1} 
%\end{table}

\section{Conclusions}\label{sec:conclusion}

We have investigated the effect of fracture rectification in graphene membranes with an array of defects of triangular shape (with different spacing) through fully atomistic reactive molecular dynamics simulations. We considered two structural models (Figure \ref{fig:uni-2-notch-2}), one with two-notches tips, to determine the preferential fracture direction propagation and once determined, with a single notch to test the fracture-like behavior.

Our results show that a robust diode-like behavior exists, thus validating that the phenomenon observed at the macroscale by Brodnik \textit{et al}~\cite{brodnik_2021} still holds at the nanoscale for graphene membranes. Other topologies and/or defects can be also exploited to guide the fracture along engineering directions. The experimental realization of the structures shown in (Figure \ref{fig:uni-2-notch-2}) is feasible today. In fact, more complex graphene kirigami structures have been already reported \cite{blees_2015}.

\section*{Acknowledgments}
This work was financed by Sao Paulo State Research Support Foundation (FAPESP). We also thank the Center for Computing in Engineering and Sciences at Unicamp for financial support through the FAPESP/CEPID Grants \#2013/08293-7 and \#2018/11352-7. 
%\newpage

\bibliographystyle{unsrtnat}
\bibliography{references}  %%% Uncomment this line and comment out the ``thebibliography'' section below to use the external .bib file (using bibtex) .

\begin{thebibliography}{43}
\providecommand{\natexlab}[1]{#1}
\providecommand{\url}[1]{\texttt{#1}}
\expandafter\ifx\csname urlstyle\endcsname\relax
  \providecommand{\doi}[1]{doi: #1}\else
  \providecommand{\doi}{doi: \begingroup \urlstyle{rm}\Url}\fi

\bibitem[Novoselov et~al.(2004)Novoselov, Geim, Morozov, Jiang, Zhang, Dubonos,
  Grigorieva, and Firsov]{novoselov_2004}
K~S Novoselov, A~K Geim, S~V Morozov, D~Jiang, Y~Zhang, S~V Dubonos, I~V
  Grigorieva, and A~A Firsov.
\newblock Electric field effect in atomically thin carbon films.
\newblock \emph{Science}, 306\penalty0 (5696):\penalty0 666--669, oct 2004.
\newblock \doi{10.1126/science.1102896}.
\newblock URL \url{http://dx.doi.org/10.1126/science.1102896}.

\bibitem[Geim and Novoselov(2007)]{geim_2007}
A~K Geim and K~S Novoselov.
\newblock The rise of graphene.
\newblock \emph{Nature Materials}, 6\penalty0 (3):\penalty0 183--191, mar 2007.
\newblock \doi{10.1038/nmat1849}.
\newblock URL \url{http://dx.doi.org/10.1038/nmat1849}.

\bibitem[Zhang et~al.(2005)Zhang, Tan, Stormer, and Kim]{zhang_2005}
Yuanbo Zhang, Yan-Wen Tan, Horst~L Stormer, and Philip Kim.
\newblock Experimental observation of the quantum hall effect and berry's phase
  in graphene.
\newblock \emph{Nature}, 438\penalty0 (7065):\penalty0 201--204, nov 2005.
\newblock \doi{10.1038/nature04235}.
\newblock URL \url{http://dx.doi.org/10.1038/nature04235}.

\bibitem[Young and Kim(2009)]{young_2009}
Andrea~F. Young and Philip Kim.
\newblock Quantum interference and klein tunnelling in graphene
  heterojunctions.
\newblock \emph{Nature physics}, 5\penalty0 (3):\penalty0 222--226, mar 2009.
\newblock ISSN 1745-2473.
\newblock \doi{10.1038/nphys1198}.
\newblock URL \url{http://www.nature.com/articles/nphys1198}.

\bibitem[Castro~Neto et~al.(2009)Castro~Neto, Guinea, Peres, Novoselov, and
  Geim]{castroneto_2009}
A.~H. Castro~Neto, F.~Guinea, N.~M.~R. Peres, K.~S. Novoselov, and A.~K. Geim.
\newblock The electronic properties of graphene.
\newblock \emph{Reviews of Modern Physics}, 81\penalty0 (1):\penalty0 109--162,
  jan 2009.
\newblock ISSN 0034-6861.
\newblock \doi{10.1103/{RevModPhys}.81.109}.
\newblock URL \url{http://link.aps.org/doi/10.1103/{RevModPhys}.81.109}.

\bibitem[Lee et~al.(2008)Lee, Wei, Kysar, and Hone]{lee_2008}
Changgu Lee, Xiaoding Wei, Jeffrey~W Kysar, and James Hone.
\newblock Measurement of the elastic properties and intrinsic strength of
  monolayer graphene.
\newblock \emph{Science}, 321\penalty0 (5887):\penalty0 385--388, jul 2008.
\newblock \doi{10.1126/science.1157996}.
\newblock URL \url{http://dx.doi.org/10.1126/science.1157996}.

\bibitem[Bizao et~al.(2017)Bizao, Botari, Perim, Pugno, and Galvao]{bizao_2017}
R.A. Bizao, T.~Botari, E.~Perim, Nicola~M. Pugno, and D.S. Galvao.
\newblock Mechanical properties and fracture patterns of graphene (graphitic)
  nanowiggles.
\newblock \emph{Carbon}, 119:\penalty0 431--437, aug 2017.
\newblock ISSN 00086223.
\newblock \doi{10.1016/j.carbon.2017.04.018}.
\newblock URL
  \url{http://linkinghub.elsevier.com/retrieve/pii/S0008622317303743}.

\bibitem[Bizao et~al.(2018)Bizao, Machado, de~Sousa, Pugno, and
  Galvao]{bizao_2018}
Rafael~A Bizao, Leonardo~D Machado, Jose~M de~Sousa, Nicola~M Pugno, and
  Douglas~S Galvao.
\newblock Scale effects on the ballistic penetration of graphene sheets.
\newblock \emph{Scientific Reports}, 8\penalty0 (1):\penalty0 6750, apr 2018.
\newblock \doi{10.1038/s41598-018-25050-2}.
\newblock URL \url{http://dx.doi.org/10.1038/s41598-018-25050-2}.

\bibitem[Fonseca and Galvao(2021)]{fonseca_2021}
Alexandre~F. Fonseca and Douglas~S. Galvao.
\newblock Graphene-based nanoscale version of da vinci's reciprocal structures.
\newblock \emph{Computational Materials Science}, 187:\penalty0 110105, feb
  2021.
\newblock ISSN 09270256.
\newblock \doi{10.1016/j.commatsci.2020.110105}.
\newblock URL
  \url{https://linkinghub.elsevier.com/retrieve/pii/S0927025620305966}.

\bibitem[Lee et~al.(2012)Lee, Yoon, and Cheong]{lee_2012}
Jae-Ung Lee, Duhee Yoon, and Hyeonsik Cheong.
\newblock Estimation of young's modulus of graphene by raman spectroscopy.
\newblock \emph{Nano Letters}, 12\penalty0 (9):\penalty0 4444--4448, sep 2012.
\newblock \doi{10.1021/nl301073q}.
\newblock URL \url{http://dx.doi.org/10.1021/nl301073q}.

\bibitem[Branicio et~al.(2016)Branicio, Vastola, Jhon, Sullivan, Shenoy, and
  Srolovitz]{branicio_2016}
Paulo~S. Branicio, Guglielmo Vastola, Mark~H. Jhon, Michael~B. Sullivan,
  Vivek~B. Shenoy, and David~J. Srolovitz.
\newblock Elastic interaction of hydrogen atoms on graphene: A multiscale
  approach from first principles to continuum elasticity.
\newblock \emph{Physical Review B}, 94\penalty0 (16):\penalty0 165420, oct
  2016.
\newblock ISSN 2469-9950.
\newblock \doi{10.1103/{PhysRevB}.94.165420}.
\newblock URL \url{https://link.aps.org/doi/10.1103/{PhysRevB}.94.165420}.

\bibitem[Zhang et~al.(2014)Zhang, Ma, Fan, Zeng, Peng, Loya, Liu, Gong, Zhang,
  Zhang, Ajayan, Zhu, and Lou]{zhang_2014}
Peng Zhang, Lulu Ma, Feifei Fan, Zhi Zeng, Cheng Peng, Phillip~E Loya, Zheng
  Liu, Yongji Gong, Jiangnan Zhang, Xingxiang Zhang, Pulickel~M Ajayan, Ting
  Zhu, and Jun Lou.
\newblock Fracture toughness of graphene.
\newblock \emph{Nature Communications}, 5:\penalty0 3782, apr 2014.
\newblock \doi{10.1038/ncomms4782}.
\newblock URL \url{http://dx.doi.org/10.1038/ncomms4782}.

\bibitem[Ashby(2011)]{ashby_2011}
Michael~F. Ashby.
\newblock \emph{Materials selection in mechanical design}.
\newblock Butterworth-Heinemann, 4th edition, 2011.
\newblock ISBN 9781856176637.
\newblock \doi{10.1016/C2009-0-25539-5}.
\newblock URL \url{https://linkinghub.elsevier.com/retrieve/pii/C20090255395}.

\bibitem[Young et~al.(2012)Young, Kinloch, Gong, and Novoselov]{young_2012}
Robert~J. Young, Ian~A. Kinloch, Lei Gong, and Kostya~S. Novoselov.
\newblock The mechanics of graphene nanocomposites: A review.
\newblock \emph{Composites science and technology}, 72\penalty0 (12):\penalty0
  1459--1476, jul 2012.
\newblock ISSN 02663538.
\newblock \doi{10.1016/j.compscitech.2012.05.005}.
\newblock URL
  \url{http://linkinghub.elsevier.com/retrieve/pii/S0266353812001789}.

\bibitem[Kim et~al.(2013{\natexlab{a}})Kim, Lee, Yeom, Shin, Kim, Cui, Kysar,
  Hone, Jung, Jeon, and Han]{kim_2013}
Youbin Kim, Jinsup Lee, Min~Sun Yeom, Jae~Won Shin, Hyungjun Kim, Yi~Cui,
  Jeffrey~W Kysar, James Hone, Yousung Jung, Seokwoo Jeon, and Seung~Min Han.
\newblock Strengthening effect of single-atomic-layer graphene in
  metal-graphene nanolayered composites.
\newblock \emph{Nature Communications}, 4:\penalty0 2114, 2013{\natexlab{a}}.
\newblock \doi{10.1038/ncomms3114}.
\newblock URL \url{http://dx.doi.org/10.1038/ncomms3114}.

\bibitem[Yang et~al.(2013)Yang, Rigdon, Huang, and Li]{yang_2013}
Yingchao Yang, William Rigdon, Xinyu Huang, and Xiaodong Li.
\newblock Enhancing graphene reinforcing potential in composites by hydrogen
  passivation induced dispersion.
\newblock \emph{Scientific Reports}, 3:\penalty0 2086, 2013.
\newblock \doi{10.1038/srep02086}.
\newblock URL \url{http://dx.doi.org/10.1038/srep02086}.

\bibitem[Callister~Jr. and Rethwisch(2018)]{callisterjr_2018}
William~D. Callister~Jr. and David~G. Rethwisch.
\newblock \emph{Materials Science and Engineering: An Introduction}.
\newblock Wiley, 10th edition, 2018.
\newblock URL
  \url{https://www.wiley.com/en-us/Materials+Science+and+Engineering\%{3A}+An+Introduction\%{2C}+10th+Edition-p-9781119405498}.

\bibitem[Sha et~al.(2015)Sha, Pei, Ding, Jiang, and Zhang]{sha_2015}
Zhen-Dong Sha, Qing-Xiang Pei, Zhiwei Ding, Jin-Wu Jiang, and Yong-Wei Zhang.
\newblock Mechanical properties and fracture behavior of single-layer
  phosphorene at finite temperatures.
\newblock \emph{Journal of physics D: Applied physics}, 48\penalty0
  (39):\penalty0 395303, oct 2015.
\newblock ISSN 0022-3727.
\newblock \doi{10.1088/0022-3727/48/39/395303}.
\newblock URL
  \url{http://stacks.iop.org/0022-3727/48/i=39/a=395303?key=crossref.757a2fb537d1a95003ef80988e7ac620}.

\bibitem[Liu et~al.(2019)Liu, Becton, Zhang, Chen, Zeng, Pidaparti, and
  Wang]{liu_2019}
Ning Liu, Matthew Becton, Liuyang Zhang, Heng Chen, Xiaowei Zeng, Ramana
  Pidaparti, and Xianqiao Wang.
\newblock A coarse-grained model for mechanical behavior of phosphorene sheets.
\newblock \emph{Physical Chemistry Chemical Physics}, 21\penalty0 (4):\penalty0
  1884--1894, jan 2019.
\newblock \doi{10.1039/c8cp06918b}.
\newblock URL \url{http://dx.doi.org/10.1039/c8cp06918b}.

\bibitem[Daniel et~al.(2004)Daniel, Sircar, Gliem, and Chaudhury]{daniel_2004}
Susan Daniel, Sanjoy Sircar, Jill Gliem, and Manoj~K Chaudhury.
\newblock Ratcheting motion of liquid drops on gradient surfaces.
\newblock \emph{Langmuir: the {ACS} Journal of Surfaces and Colloids},
  20\penalty0 (10):\penalty0 4085--4092, may 2004.
\newblock \doi{10.1021/la036221a}.
\newblock URL \url{http://dx.doi.org/10.1021/la036221a}.

\bibitem[Malvadkar et~al.(2010)Malvadkar, Hancock, Sekeroglu, Dressick, and
  Demirel]{malvadkar_2010}
Niranjan~A Malvadkar, Matthew~J Hancock, Koray Sekeroglu, Walter~J Dressick,
  and Melik~C Demirel.
\newblock An engineered anisotropic nanofilm with unidirectional wetting
  properties.
\newblock \emph{Nature Materials}, 9\penalty0 (12):\penalty0 1023--1028, dec
  2010.
\newblock \doi{10.1038/nmat2864}.
\newblock URL \url{http://dx.doi.org/10.1038/nmat2864}.

\bibitem[Hancock et~al.(2012)Hancock, Sekeroglu, and Demirel]{hancock_2012}
Matthew~J Hancock, Koray Sekeroglu, and Melik~C Demirel.
\newblock Bioinspired directional surfaces for adhesion, wetting and transport.
\newblock \emph{Advanced functional materials}, 22\penalty0 (11):\penalty0
  2223--2234, jun 2012.
\newblock \doi{10.1002/adfm.201103017}.
\newblock URL \url{http://dx.doi.org/10.1002/adfm.201103017}.

\bibitem[Zambrano et~al.(2018)Zambrano, Gallardo, Polania, Rodríguez, and
  Coronado]{zambrano_2018}
O.~A. Zambrano, K.~F. Gallardo, D.~M. Polania, S.~A. Rodríguez, and J.~J.
  Coronado.
\newblock The role of the counterbody's oxide on the wear behavior of {H\SS}
  and hi-cr.
\newblock \emph{Tribology letters}, 66\penalty0 (1):\penalty0 1, mar 2018.
\newblock ISSN 1023-8883.
\newblock \doi{10.1007/s11249-017-0954-1}.
\newblock URL \url{http://link.springer.com/10.1007/s11249-017-0954-1}.

\bibitem[Xia et~al.(2012)Xia, Ponson, Ravichandran, and Bhattacharya]{xia_2012}
S~Xia, L~Ponson, G~Ravichandran, and K~Bhattacharya.
\newblock Toughening and asymmetry in peeling of heterogeneous adhesives.
\newblock \emph{Physical Review Letters}, 108\penalty0 (19):\penalty0 196101,
  may 2012.
\newblock \doi{10.1103/{PhysRevLett}.108.196101}.
\newblock URL \url{http://dx.doi.org/10.1103/{PhysRevLett}.108.196101}.

\bibitem[Hossain et~al.(2014)Hossain, Hsueh, Bourdin, and
  Bhattacharya]{hossain_2014}
M.Z. Hossain, C.-J. Hsueh, B.~Bourdin, and K.~Bhattacharya.
\newblock Effective toughness of heterogeneous media.
\newblock \emph{Journal of the mechanics and physics of solids}, 71:\penalty0
  15--32, nov 2014.
\newblock ISSN 00225096.
\newblock \doi{10.1016/j.jmps.2014.06.002}.
\newblock URL
  \url{https://linkinghub.elsevier.com/retrieve/pii/S0022509614001215}.

\bibitem[Brodnik et~al.(2021)Brodnik, Brach, Long, Ravichandran, Bourdin,
  Faber, and Bhattacharya]{brodnik_2021}
N~R Brodnik, S~Brach, C~M Long, G~Ravichandran, B~Bourdin, K~T Faber, and
  K~Bhattacharya.
\newblock Fracture diodes: directional asymmetry of fracture toughness.
\newblock \emph{Physical Review Letters}, 126\penalty0 (2):\penalty0 025503,
  jan 2021.
\newblock \doi{10.1103/{PhysRevLett}.126.025503}.
\newblock URL \url{http://dx.doi.org/10.1103/{PhysRevLett}.126.025503}.

\bibitem[Conway et~al.(2021)Conway, Kunka, White, Pataky, and
  Boyce]{conway_2021}
Kaitlynn~M. Conway, Cody Kunka, Benjamin~C. White, Garrett~J. Pataky, and
  Brad~L. Boyce.
\newblock Increasing fracture toughness via architected porosity.
\newblock \emph{Materials \& Design}, 205:\penalty0 109696, jul 2021.
\newblock ISSN 02641275.
\newblock \doi{10.1016/j.matdes.2021.109696}.
\newblock URL
  \url{https://linkinghub.elsevier.com/retrieve/pii/S0264127521002483}.

\bibitem[Dragoman et~al.(2010)Dragoman, Dragoman, and Plana]{dragoman_2010}
D.~Dragoman, M.~Dragoman, and R.~Plana.
\newblock Graphene-based ultrafast diode.
\newblock \emph{Journal of applied physics}, 108\penalty0 (8):\penalty0 084316,
  oct 2010.
\newblock ISSN 0021-8979.
\newblock \doi{10.1063/1.3501051}.
\newblock URL \url{http://aip.scitation.org/doi/10.1063/1.3501051}.

\bibitem[Di~Bartolomeo(2016)]{dibartolomeo_2016}
Antonio Di~Bartolomeo.
\newblock Graphene schottky diodes: An experimental review of the rectifying
  graphene/semiconductor heterojunction.
\newblock \emph{Physics Reports}, 606:\penalty0 1--58, jan 2016.
\newblock ISSN 03701573.
\newblock \doi{10.1016/j.physrep.2015.10.003}.
\newblock URL
  \url{http://linkinghub.elsevier.com/retrieve/pii/S0370157315004354}.

\bibitem[Wang et~al.(2021)Wang, Hemmetter, Uzlu, Saeed, Hamed, Kataria, Negra,
  Neumaier, and Lemme]{wang_2021}
Zhenxing Wang, Andreas Hemmetter, Burkay Uzlu, Mohamed Saeed, Ahmed Hamed,
  Satender Kataria, Renato Negra, Daniel Neumaier, and Max~C. Lemme.
\newblock Graphene in {2D}/{3D} heterostructure diodes for high performance
  electronics and optoelectronics.
\newblock \emph{Advanced Electronic Materials}, 7\penalty0 (7):\penalty0
  2001210, jul 2021.
\newblock ISSN 2199-{160X}.
\newblock \doi{10.1002/aelm.202001210}.
\newblock URL \url{https://onlinelibrary.wiley.com/doi/10.1002/aelm.202001210}.

\bibitem[Kim et~al.(2013{\natexlab{b}})Kim, Lee, {McEvoy}, Yim, and
  Duesberg]{kim_2013a}
Hye-Young Kim, Kangho Lee, Niall {McEvoy}, Chanyoung Yim, and Georg~S Duesberg.
\newblock Chemically modulated graphene diodes.
\newblock \emph{Nano Letters}, 13\penalty0 (5):\penalty0 2182--2188, may
  2013{\natexlab{b}}.
\newblock \doi{10.1021/nl400674k}.
\newblock URL \url{http://dx.doi.org/10.1021/nl400674k}.

\bibitem[Li et~al.(2016)Li, Lin, Lin, Xu, Wang, Zhang, Zhong, Xu, Wu, and
  Fang]{li_2016}
Xiaoqiang Li, Shisheng Lin, Xing Lin, Zhijuan Xu, Peng Wang, Shengjiao Zhang,
  Huikai Zhong, Wenli Xu, Zhiqian Wu, and Wei Fang.
\newblock Graphene/h-{BN}/{GaAs} sandwich diode as solar cell and
  photodetector.
\newblock \emph{Optics Express}, 24\penalty0 (1):\penalty0 134--145, jan 2016.
\newblock \doi{10.1364/{OE}.24.000134}.
\newblock URL \url{http://dx.doi.org/10.1364/{OE}.24.000134}.

\bibitem[Wang et~al.(2017)Wang, Hu, Takahashi, Zhang, Takamatsu, and
  Chen]{wang_2017}
Haidong Wang, Shiqian Hu, Koji Takahashi, Xing Zhang, Hiroshi Takamatsu, and
  Jie Chen.
\newblock Experimental study of thermal rectification in suspended monolayer
  graphene.
\newblock \emph{Nature Communications}, 8:\penalty0 15843, jun 2017.
\newblock \doi{10.1038/ncomms15843}.
\newblock URL \url{http://dx.doi.org/10.1038/ncomms15843}.

\bibitem[Liu et~al.(2021)Liu, Muruganathan, Feng, Ogawa, Morita, Liu, Guo,
  Schmidt, and Mizuta]{liu_2021}
Fayong Liu, Manoharan Muruganathan, Yu~Feng, Shinichi Ogawa, Yukinori Morita,
  Chunmeng Liu, Jiayu Guo, Marek Schmidt, and Hiroshi Mizuta.
\newblock Thermal rectification on asymmetric suspended graphene nanomesh
  devices.
\newblock \emph{Nano Futures}, 5\penalty0 (4):\penalty0 045002, dec 2021.
\newblock ISSN 2399-1984.
\newblock \doi{10.1088/2399-1984/ac36b5}.
\newblock URL
  \url{https://iopscience.iop.org/article/10.1088/2399-1984/ac36b5}.

\bibitem[Chen et~al.(2020)Chen, Pang, Chen, Du, and Chen]{chen_2020}
Xue-Kun Chen, Min Pang, Tong Chen, Dan Du, and Ke-Qiu Chen.
\newblock Thermal rectification in asymmetric graphene/hexagonal boron nitride
  van der waals heterostructures.
\newblock \emph{{ACS} Applied Materials \& Interfaces}, 12\penalty0
  (13):\penalty0 15517--15526, apr 2020.
\newblock \doi{10.1021/acsami.9b22498}.
\newblock URL \url{http://dx.doi.org/10.1021/acsami.9b22498}.

\bibitem[Melis et~al.(2015)Melis, Barbarino, and Colombo]{melis_2015}
Claudio Melis, Giuliana Barbarino, and Luciano Colombo.
\newblock Exploiting hydrogenation for thermal rectification in graphene
  nanoribbons.
\newblock \emph{Physical Review B}, 92\penalty0 (24):\penalty0 245408, dec
  2015.
\newblock ISSN 1098-0121.
\newblock \doi{10.1103/{PhysRevB}.92.245408}.
\newblock URL \url{https://link.aps.org/doi/10.1103/{PhysRevB}.92.245408}.

\bibitem[Wang et~al.(2014)Wang, Vallabhaneni, Hu, Qiu, Chen, and
  Ruan]{wang_2014}
Yan Wang, Ajit Vallabhaneni, Jiuning Hu, Bo~Qiu, Yong~P Chen, and Xiulin Ruan.
\newblock Phonon lateral confinement enables thermal rectification in
  asymmetric single-material nanostructures.
\newblock \emph{Nano Letters}, 14\penalty0 (2):\penalty0 592--596, feb 2014.
\newblock \doi{10.1021/nl403773f}.
\newblock URL \url{http://dx.doi.org/10.1021/nl403773f}.

\bibitem[Plimpton(1995)]{plimpton_1995}
Steve Plimpton.
\newblock Fast parallel algorithms for short-range molecular dynamics.
\newblock \emph{Journal of Computational Physics}, 117\penalty0 (1):\penalty0
  1--19, mar 1995.
\newblock ISSN 00219991.
\newblock \doi{10.1006/jcph.1995.1039}.
\newblock URL
  \url{http://linkinghub.elsevier.com/retrieve/pii/{S002199918571039X}}.

\bibitem[Stuart et~al.(2000)Stuart, Tutein, and Harrison]{stuart_2000}
Steven~J. Stuart, Alan~B. Tutein, and Judith~A. Harrison.
\newblock A reactive potential for hydrocarbons with intermolecular
  interactions.
\newblock \emph{The Journal of Chemical Physics}, 112\penalty0 (14):\penalty0
  6472--6486, apr 2000.
\newblock ISSN 0021-9606.
\newblock \doi{10.1063/1.481208}.
\newblock URL \url{http://aip.scitation.org/doi/10.1063/1.481208}.

\bibitem[Shenderova et~al.(2000)Shenderova, Brenner, Omeltchenko, Su, and
  Yang]{shenderova_2000}
O.~A. Shenderova, D.~W. Brenner, A.~Omeltchenko, X.~Su, and L.~H. Yang.
\newblock Atomistic modeling of the fracture of polycrystalline diamond.
\newblock \emph{Physical Review B}, 61\penalty0 (6):\penalty0 3877--3888, feb
  2000.
\newblock ISSN 0163-1829.
\newblock \doi{10.1103/{PhysRevB}.61.3877}.
\newblock URL \url{https://link.aps.org/doi/10.1103/{PhysRevB}.61.3877}.

\bibitem[Mises(1913)]{mises_1913}
R.~v. Mises.
\newblock Mechanik der festen körper in plastisch-deformablen zustand.
\newblock \emph{Math.-phys. Klasse}, 4:\penalty0 582--592, 1913.
\newblock URL \url{https://eudml.org/doc/58894}.

\bibitem[Stukowski(2010)]{stukowski_2010}
Alexander Stukowski.
\newblock Visualization and analysis of atomistic simulation data with
  {OVITO}–the open visualization tool.
\newblock \emph{Modelling and Simulation in Materials Science and Engineering},
  18\penalty0 (1):\penalty0 015012, jan 2010.
\newblock ISSN 0965-0393.
\newblock \doi{10.1088/0965-0393/18/1/015012}.
\newblock URL
  \url{http://stacks.iop.org/0965-0393/18/i=1/a=015012?key=crossref.6895e2c3bb522d1563fb2e2fe9f22789}.

\bibitem[Blees et~al.(2015)Blees, Barnard, Rose, Roberts, {McGill}, Huang,
  Ruyack, Kevek, Kobrin, Muller, and {McEuen}]{blees_2015}
Melina~K Blees, Arthur~W Barnard, Peter~A Rose, Samantha~P Roberts, Kathryn~L
  {McGill}, Pinshane~Y Huang, Alexander~R Ruyack, Joshua~W Kevek, Bryce Kobrin,
  David~A Muller, and Paul~L {McEuen}.
\newblock Graphene kirigami.
\newblock \emph{Nature}, 524\penalty0 (7564):\penalty0 204--207, aug 2015.
\newblock \doi{10.1038/nature14588}.
\newblock URL \url{http://dx.doi.org/10.1038/nature14588}.

\end{thebibliography}





% \end{thebibliography}

\end{document}